\documentclass[aps,twocolumn,apl,superscriptaddress,showpacs,longbibliography]{revtex4-1}

\usepackage{graphicx}
\usepackage{dcolumn}
\usepackage{bm}
\usepackage{upgreek}
\usepackage{paralist}

\newcommand{\EECS}{\affiliation{Department of Electrical Engineering and Computer Science, Massachusetts Institute of Technology, Cambridge MA 02139}}

\begin{document}

\title{A Tunable Waveguide-Coupled Cavity Design for Efficient Spin-Photon Interfaces in Photonic Integrated Circuits}
\author{Sara L. Mouradian}
\email{smouradi@mit.edu}
\author{Dirk Englund}
\email{englund@mit.edu}
\EECS
\begin{abstract}

A solid state emitter coupled to a photonic crystal cavity exhibits increased photon emission into a single frequency mode. However, current designs for photonic crystal cavities coupled to quantum emitters have three main problems: emitters are placed near surfaces that can degrade their optical properties, the cavity fluorescence cannot be collected into a single useful mode for further routing, and post-fabrication tuning is not currently possible in a stable and reversible manner for each node individually. In this paper, we introduce a hybrid cavity design with minimal fabrication of the host material that keeps the emitter $\geq100\,$nm from all surfaces. This cavity has an unloaded quality factor ($Q$) larger than $1\times10^6$ and a loaded $Q$ of $5.5\times10^4$ with more than 75\% of the emission coupled directly into an underlying photonic integrated circuit built from a convenient material that provides low loss waveguides. Finally this design can be actively and reversibly tuned onto resonance with the emitter, allowing tuning over more than 10 times the cavity linewidth while maintaining $\geq50$\% of the $Q$ factor with no effects to other cavities on the same chip. 

\end{abstract}

\maketitle

Solid state quantum emitters, such as defects in diamond or quantum dots in III-V materials, have been demonstrated as high quality single photon sources and quantum memories. The entanglement rate between quantum repeaters in a quantum communication network, or between quantum bits (qubits) in a distributed quantum computation platform scales with the square of the collection rate of indistinguishable photons. Thus, the cavity enhancement of the transition of interest can significantly increase secure communication rates as well as the size of an entangled network. Photonic crystal (PhC) cavities are an attractive choice for solid state qubits as they provide high quality factors ($Q$) with wavelength-scale mode volumes ($V$) with direct patterning of the host material. Cavity enhancement of quantum dots in III-V materials has enabled strong light-matter coupling~\cite{englund2007controlling} and record rates of indistinguishable photons~\cite{santori2002indistinguishable,somaschi2016near}. Recent advances in diamond nanofabrication~\cite{Schroder2016_Review} have enabled enhancement of optically active defect centers in diamond. Silicon vacancy centers coupled to PhC cavities increased emission rates and mediated entanglement between multiple centers~\cite{riedrich2014deterministic,sipahigil2016single}. The zero phonon line (ZPL) transition of the negatively charged nitrogen vacancy (NV) center in diamond has also been enhanced by PhC cavities~\cite{li2015coherent,wolters2010enhancement,englund2010deterministic,2014Igal_APL,burek2014high, schukraft2016imp,faraon2011resonant,faraon2012coupling}.

\begin{figure}[htp]
\begin{center}
\includegraphics[width=3.5in]{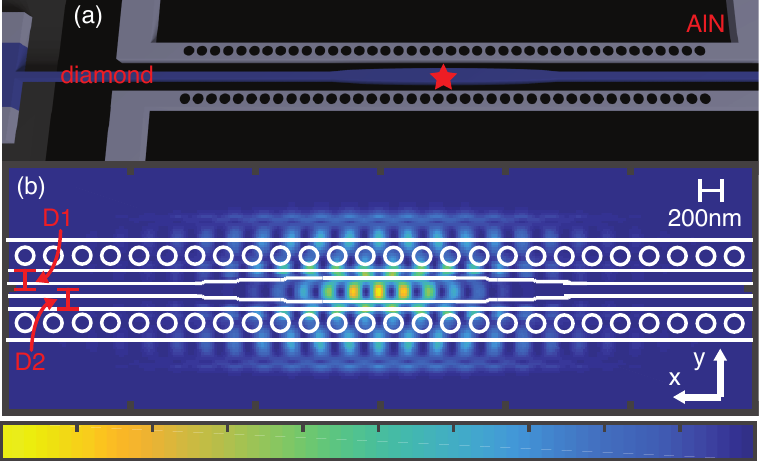}
\end{center}
\caption{(a) Structure of the hybrid cavity. Two suspended AlN beams (gray) periodically patterned with holes sit on either side of a width-modulated diamond (blue). (b) Unloaded cavity mode profile $(|E|^2)$. }
\label{struct}
\end{figure}

There are currently three main challenges limiting progress in scaling to multiple coupled emitter-cavity systems. (1) Nanofabrication of PhC cavities can introduce defects in the crystal and on the surface due to the ion bombardment that is necessary to etch the host material. These defects trap charges in unstable configurations that lead to pure dephasing and spectral diffusion that in turn lead to a decrease in entanglement rate due to a loss of indistinguishability between photons emitted by distinct emitters. (2) In order to improve the entanglement rate, the cavity design must not only enhance the emission of the desired transition via a high Q mode at the transition frequency, but also increase the collection efficiency of that transition. Thus the loss pathways of the cavity must be engineered to funnel the enhanced light into a single useable mode that is indistinguishable with other nodes. (3) Finally, fabrication inconsistencies across a single chip cause a spread in the resonant frequencies of the final devices that can severely decrease the enhancement of the desired transition. 

In this letter, we present a hybrid cavity design that addresses all three of these challenges: First, the emitter's host material is minimally patterned, and the emitter remains 100\,nm from every surface. The periodic change in the dielectric environment needed to produce a band gap is provided by the patterning of a second material. Second, $\>75$\% of the mode is coupled into a single waveguide mode. Finally, the frequency is tunable over 10 linewidths while maintaining a $Q$ factor within 50\% of the maximum value.

While the design presented in this paper can be easily modified to enhance other solid state emitters, or connect with other PIC systems, we focus on a design to enhance the NV center's ZPL in a diamond node integrated into an Aluminum Nitride (AlN) PIC. 

The NV center has exceptional spin properties for a solid state quantum emitter, with second-scale electron spin coherence times~\cite{2012Maurer,2013.NComm.Bar-Gill-} that can be optically initialized, manipulated, and measured~\cite{2010Neumann_Science_SSRO}, and can be mapped onto nearby auxiliary nuclear memories~\cite{2007Dutt_Sci_register} for increased coherence times. The long coherence time and ease of state measurement and manipulation enables entanglement generation and state teleportation between two spatially separate quantum memories~\cite{2014Pfaff_Sci_teleport,2013Bernien_Nat_entangle,2016Hensen}. However, the reported entanglement rates remain well below even the nuclear spin decoherence rate, eliminating the possibility of creating entanglement over three or more nodes. This low entanglement rate is currently limited by the small probability of collecting a photon coherent with the spin state into the frequency and spatial mode needed for entanglement due to low collection efficiency, spectral diffusion, and phonon interactions present even at cryogenic temperatures. Thus, cavity enhancement is essential to scale entanglement from two NV centers to three or more. AlN is chosen to create the optical band-gap at the cavity region, and as the backbone PIC material as it has a wide direct band-gap (6.015\,eV) that allows low-loss single-mode operation at the NV ZPL wavelength (637\,nm). Moreover, nanoelectromechanical functionality~\cite{karabalin2009piezoelectric,sinha2009piezoelectric}, and frequency conversion~\cite{guo2016chip,zou2016cavity} have been demonstrated that make it a promising platform for reconfigurable quantum circuits that can interface with fibers for long-distance communication. 

As shown in Figure~\ref{struct}(a), the cavity consists of an underlying AlN PIC---suspended at the cavity region---with a height $H/\lambda_{\text{NV}}=0.345$ where $\lambda_{\text{NV}}=637$\,nm is the free-space ZPL wavelength. Two parallel AlN beams (with width $W/\lambda_{\text{NV}} = 0.383$) are patterned with periodic holes with constant spacing ($a/\lambda_{\text{NV}}=0.356$) and constant radius ($r/a=0.295$). A diamond slab with height $H/\lambda_{\text{NV}}=0.345$) containing a single NV center is placed between the two AlN beams, centered at a distance $D_1=D_2=0.235\lambda_{\text{NV}}$ from each AlN beam. The width of the suspended diamond membrane increases parabolically from $W_1/\lambda_{\text{NV}}=0.157$ to $W_2/\lambda_{\text{NV}}=0.314$ over 7 periods. The periodic patterning of the two AlN waveguides creates a bandgap in the $x$ direction, while the increased width of the diamond membrane at the center of the structure increases the effective refractive index in a localized spot in the otherwise periodic structure, pulling the air bands into the diamond. Finite difference time domain (FDTD) simulations show a radiation-limited $Q$ factor of $>1$ million with 50 holes on either side of the center of the cavity. Figure~\ref{struct}b shows the energy distribution $(|E|^2)$ in the cavity with a clear maximum in the diamond due to the high index ($n=2.4$).

FDTD simulations reveal a mode volume of $V \approx 10 (\lambda_{NV}/n)^3$. Due to the large transverse extent of the hybrid structure, this mode volume is larger than other nanobeam cavity designs, which provide mode volumes on the order of $(\lambda/n)^3$. However, the high $Q$ ensures a large Purcell enhancement. Additionally, the bulk of the patterning is done in the AlN, leaving the NV center 100\,nm from every fabricated surface, while still maintaining a high $Q$ with a mode volume on the order of the resonant wavelength. This is essential because the spectral diffusion of solid state emitters increases with proximity to fabricated surfaces~\cite{faraon2012coupling,2015_PRX_Mouradian}, and previous work in bulk diamond has demonstrated lifetime limited linewidths of NVs 100\,nm from the surface~\cite{chu2014coherent}.

\begin{figure}
\begin{center}
\includegraphics[width=3.5in]{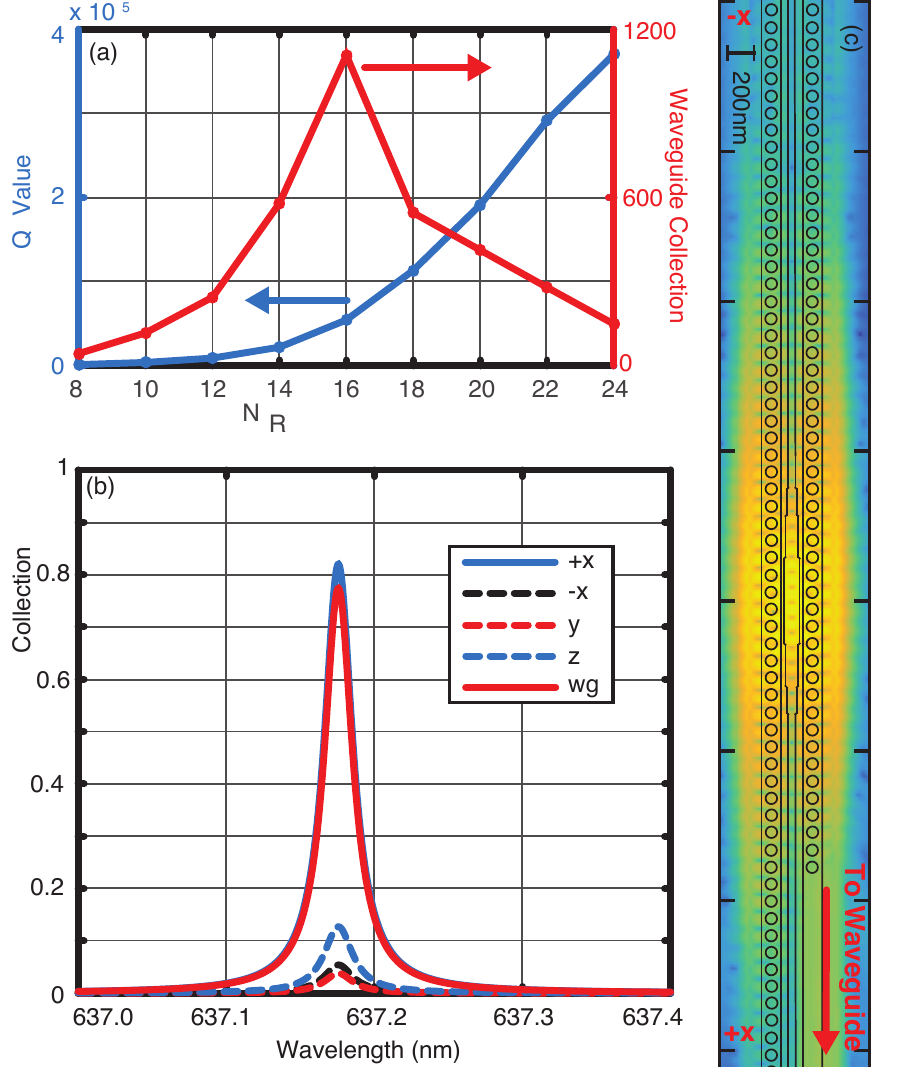}
\end{center}
\caption{(a) The enhancement of collection into the waveguide mode vs the number of holes leading to the loading waveguide (b) The collection spectrum into the waveguide mode, as well as the losses into the y and z planes at $N_R=16$. (c) Loaded cavity mode profile $(|E|^2)$. }
\label{wgCouple}
\end{figure}

To efficiently collect the NV emission into the PIC, the cavity must be coupled to the waveguide. The unloaded design described above has a high $Q$, but the emitted light cannot be efficiently collected into the waveguide as it radiates mainly into the $z$-plane. Thus, despite the large enhancement due to the high $Q/V$, there is only a marginal increase of light collected into a single AlN waveguide mode over a simpler, broadband architecture of an NV coupled to a single diamond waveguide mode that is adiabatically tapered to transfer the optical mode to the underlying PIC mode~\cite{2015_PRX_Mouradian}. The cavity is loaded into a single waveguide mode by reducing the number of holes on one of the four sides of the cavity, as seen in Fig.~\ref{wgCouple}(a). In general, the fraction of emission into the waveguide can be estimated from the ratio $F=Q_l/(Q_l+Q_i)$ where $Q_i$ is the intrinsic Q factor, and $Q_l$ is the loaded $Q$ factor. However, modifying the cavity geometry can also increase scattering into other loss pathway. Moreover, the figure of merit of interest is the overall collection enhancement, which is a function of $Q$ and $F$. Therefore, we performed detailed FDTD simulations to measure the collection enhancement into the waveguide port. As seen in Fig.~\ref{wgCouple}(a), the collection enhancement is maximized when the number of holes leading to the output waveguide is $N_R=16$. At this configuration, the loaded $Q$ is 55,000, with 75\% of the light coupled into the waveguide as calculated with FDTD simulations. The collection spectra into the waveguide, as well as into the $y$ and $z$ planes is shown in Fig.~\ref{wgCouple}b. Fig.~\ref{wgCouple}c shows the mode profile (log$(|E|^2)$) of the loaded cavity, with clear coupling into the bottom right waveguide. 

\begin{figure}
\begin{center}
\includegraphics[width=3in]{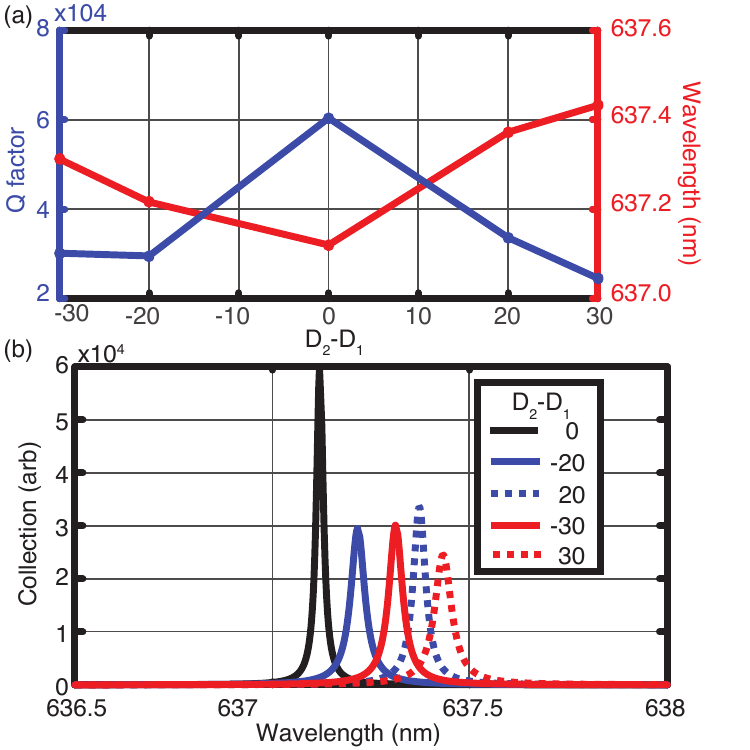}
\end{center}
\caption{(a) Resonant wavelength and $Q$ factor as the two AlN DBRs are displaced. (b) Cavity spectrum for different positions of the AlN beams. }
\label{tuning}
\end{figure}

For efficient high-fidelity entanglement between two NV centers, they both must emit into the same Fourier-limited frequency mode. The ZPL transition frequency of 2 or more NV centers can be tuned to overlap via the Stark effect~\cite{tamarat2006stark}. However, the cavities must also be at the same frequencies. Unfortunately, fabrication imperfections across a single chip can cause inhomogeneities in the resonant frequencies. Therefore, it is necessary to tune the frequency of the cavity post-fabrication without a significant drop in the quality factor of the cavity. Previous works have demonstrated cavity tuning in other material systems. For instance, the fast and reversible tuning of PhC cavities in semiconductors such as Si and GaAs has been demonstrated via injection of free carriers, either electrically~\cite{fushman2007ultra} or via two photon absorption~\cite{tanabe2005all}. However, PN junctions are notoriously difficult to make in diamond and doping the diamond may have deleterious effects on the NV center. Previous cavity results in diamond~\cite{li2015coherent,sipahigil2016single} have used gas adsorption and sublimation to tune a cavity's resonance frequency, but this approach affects all cavities on the chip simultaneously. 
 Works in other material systems employed nano-electrical-mechanical systems (NEMS) to couple two cavity modes~\cite{chew2010plane,frank2010programmable} or modify the evanescent field of the cavity mode to induce wavelength shifts~\cite{chew2011enhanced,tian2012tuning,chew2010dynamic}. This is not currently feasible in an all-diamond system as large-scale free-standing membranes are not widely available, and moving the diamond would cause strain which would shift the NV center's optical transitions~\cite{batalov2009low}. In the hybrid design presented in this paper, the AlN layer can provide the needed NEMS functionality using electrostatic or piezoelectric~\cite{karabalin2009piezoelectric,sinha2009piezoelectric} actuation. FDTD simulations show that displacing the two patterned AlN beams changes the cavity's effective refractive index, and thus the resonant frequency. The tuning simulations were performed on the loaded cavity. Moving the waveguide-coupled AlN beam a certain distance ($D_1$) provides different tuning than moving the non-coupled beam the same distance ($D_2=D_1$). Figure~\ref{tuning}a summarizes the results showing the effect of beam displacements on resonant frequency and $Q$, plotted against ($D_2-D_1$). Figure~\ref{tuning}b demonstrates that the cavity can be tuned over 10 linewidths while maintaining $Q\geq Q_{\text{max}}$. Finally, if such a high loaded Q is not necessary, lower Q designs can be tuned farther without sacrificing $Q$. 

Finally, this hybrid cavity design is compatible with the scalable "pick-and-place" assembly of a photonic backbone populated with pre-selected diamond nodes~\cite{2015_PRX_Mouradian}. The probability of creating a single optimally positioned and aligned NV center remains at $\sim$1-3\% even with apertured nitrogen implantation~\cite{schukraft2016imp,scarabelli2016nanoscale,bayn2015generation} due to the stochastic process of NV creation. In this hybrid system however, the diamond nodes can be pre-screened and only the best integrated into the PhC cavity, leading to a high yield in the final system. 

In conclusion, we introduced a novel hybrid cavity design that enhances the collection of photons from a transition of an embedded quantum emitter that is at the resonant frequency of the cavity. This is achieved by loading a high-$Q$ cavity directly into a single mode PIC network. Moreover, the emitter remains $\geq$100\,nm away from every surface, reducing spectral diffusion due to trapped charges on the surface. Finally, NEMS actuation permits the reversible tuning of the resonance frequency. The implementation of this design across a multi-node PIC will enable efficient multi-qubit entanglement across the PIC. 

\begin{acknowledgments}
This research was supported in part by the Army Research Laboratory Center for Distributed Quantum Information (CDQI). S.M. was supported by IQUISE. We thank Mikkel Heuck and Tim Schr\"{o}der for valuable discussions
\end{acknowledgments}

\bibliography{2016_TunableCav_ArXiv,revtex-custm}

\end{document}